\documentstyle[epsfig]{article}
\begin{document}

\noindent
{\Large\bf Local Pulsars; A note on the Birth-Velocity 
Distribution\footnote{This paper is an extended analysis of the talk 
presented by AB at the Raman Research Institute on 20th February, 1996.}}

\vspace*{0.5cm}

\noindent
{\large\bf A. Blaauw$^2$ \& R. Ramachandran$^{3}$} \\ 
{\small $^2$ Department of Astronomy, Kapteyn Institute, P.O.Box 800, 9700 AV 
Groningen, The Netherlands.} \\ 
\noindent
{\small $^3$ Astronimical Institute ``Anton Pannekoek'', Kruislaan 403, 1098 SJ
Amsterdam, The Netherlands.} \\

\vspace*{0.5cm}


\noindent
{\bf Abstract.} We explore a simple model for the representation of the 
observed distributions of the motions, and the characteristic ages of the 
local population of pulsars. The principal difference from earlier models 
is the introduction of a unique value, $S$, for the kick velocity with 
which pulsars are born. We consider separately the proper motion components 
in galactic longitude and latitude, and find that the distributions of the 
velocity components parallel and perpendicular to the galactic plane are 
represented satisfactorily by $S=200$ km/sec, and leave no room for a 
significant fraction of much higher velocities. The successful proposition 
of a unique value for the kick velocity may provide an interesting tool in 
attempts to understand the physical process leading to the expulsion of 
the neutron star.

\section{Introduction; the Model used}
\label{sec-intro}
The early proper motion measurements have indicated that pulsars, on the 
average, have spatial velocities of the order of a few hundred km/sec 
(Lyne, Anderson \& Salter 1982). Subsequent measurements, over the past 15 
years have provided evidence for speeds as high as a thousand km/sec (Bailes 
{\it et al.} 1990; Fomalont {\it et al.} 1992; Harrison {\it et al.} 1993). 
Many models have been suggested to explain the spatial velocities of 
pulsars. Harrison \& Tadimaru (1975) proposed the ``rocket'' theory which 
essentially stated that the pulsar is accelerated along the magnetic axis 
due to the radiation reaction soon after it was born. However,  
subsequent observations have failed to confirm the prediction of the model, 
that there is an alignment between the directions of proper motion and the 
magnetic axis. Gott, Gunn \& Ostriker (1970) proposed that pulsars derive 
their spatial velocities from their progenitor binary systems, when the 
binary disrupts due to the heavy mass loss during the supernova explosion 
(Blaauw 1961). Asymmetric supernova explosions were proposed by Shklowskii 
(1970). No fully satisfactory theory has been proposed yet.

\begin{figure}
\epsfig{file=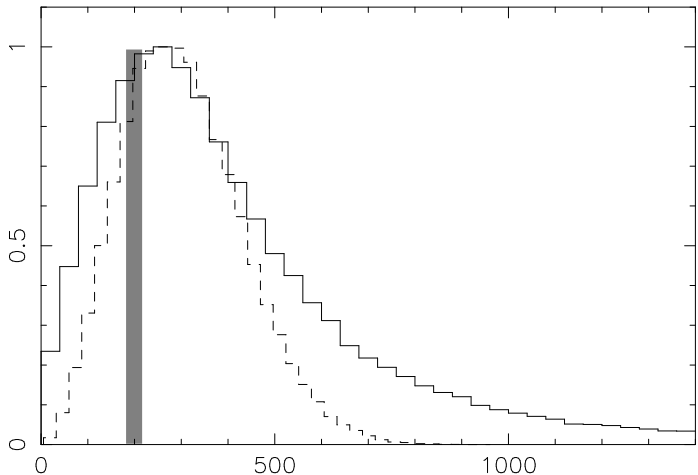,height=7cm}
\caption[]{The birth speed distribution of pulsars as predicted by Lyne \&
Lorimer (1993) (solid line) and Hansen \& Phinney (1997) (dash line), and 
as explored in the present paper (gray bar).}
\label{fig:llhp}
\end{figure}

Clearly, knowledge of the shape of the birth-velocity distribution is of
basic importance for any theory that attempts to clarify the mechanism that
causes the birth-velocities. The present paper is concerned with this
problem. Among the most recent hypotheses for the birth-speed distribution
are those by Lyne \& Lorimer (1993) and Hansen \& Phinney (1997), shown in
figure~\ref{fig:llhp}. Hartman (1997) finds that a velocity distribution
that has many more pulsars at low velocities than that of Lyne \& Lorimer
also describes the observations quite well. 

Whereas these distributions are characterised by maximum density in
velocity-space near zero-velocity, we assume
all pulsars to have been ejected from the parent body with the same birth
speed -- henceforth denoted by $S$ -- isotropically distributed.
Accordingly, in velocity space, velocity vector points define a spherical
surface with radius $S$. We consider such an alternative worth exploring
because it cannot be excluded at this stage of pulsar physics, that the
birth (kick) velocities are spread around a (physically) preferred value
rather different from zero. With regard to the space distribution of the
progenitors, we assume these to be located in a layer of zero thickness at
$z=0$. This assumption is rather harmless because it is generally agreed
that pulsars originate from the massive stars in the Galaxy which are
confined to a layer the thickness of which is very small as compared to the
scale-height of the $z$--distribution of the pulsars. Finally we assume 
an upper limit to the age, $T$, and we do not consider for this analysis 
pulsars with ages greater than $T$. Accordingly, the quantity to be 
solved for will be only $S$. Our analysis also differs from previous ones
in that we separately consider the velocity components parallel and
perpendicular to the galactic plane.

Naturally, the assumed uniqueness of the birth speed cannot but be 
an approximation of the true situation, for even if there is uniqueness in
the kick velocities with which the pulsars leave the parent body, these are
vectorially added to the space motion of the parent body, i.e., in the case of
components of binary systems, the orbital velocity and the systemic
velocity. This implies that the birth-speed distribution is not a
delta function but exhibits a certain width. However, since the birth speed
we arrive at, $S=200$ km/sec, generally dominates over these vector
additions, the simplified model remains a good first approximation.

In section \ref{sec-analsec} we give analytical expressions for the expected 
distributions, and demonstrate why the pulsar velocities parallel to the
plabe provides a direct way to estimate the velocity $S$. The observed
pulsar population suffers from severe selection effects, which must be
taken into account when comparing theoretical and observed distributions.
We have done this through a detailed Monte Carlo simulations described in
section \ref{sec-simulations}.

We compare our model with observations for the cylindrical volume of
space defined by $|z|<1$ kpc and projected distances $d\cos b$ smaller
than 2 kpc only. Within this volume we assume steady state for velocity
and space distribution.

\section{Definitions and analytical relations}
\label{sec-analsec}
We denote by,

\noindent
$S$,~~~~ the (unique) three dimensional speed at birth of the neutron star. 
In our analysis this will be the velocity with respect to the local standard 
of rest at the position of the Sun. 

\noindent
$T$,~~~~ the (adopted) upper limit to the pulsar age; only those pulsars
with characteristic ages less than $T$ are considered for this analysis.

\noindent
$z$,~~~~ the distance from the galactic plane,

\noindent
$K(z)$,~~~~ the acceleration perpendicular to the plane due to the galactic 
gravitational field; $K$ will be counted positive in the direction towards
the plane,

\noindent
$d$,~~~~the distance from the Sun,

\noindent
$w$,~~~~the velocity component  perpendicular to the galactic plane,

\noindent
$w_{0}$,~~~~the value of $w$ at birth,

\noindent
$v$,~~~~the velocity component (parallel to the plane) corresponding to
the proper motion component in galactic longitude,

\noindent
$G(S)$,~~~~the number of neutron stars generated per unit time per unit
surface at $z=0$,

\noindent
$G_{0}(w_{0})\;{\rm d}w_0$,~~~~ the number among these in the interval
$w_{0}$, $w_{0}+{\rm d}w_0$. Since the velocities at birth are
distributed isotropically, this is a flat distribution:
\begin{equation}
G_{0}(w_{0})\;{\rm d}w_0 = G(S) \;{\rm d}w_0/S = C\;{\rm d}w_0,
\end{equation}

\noindent
$G_{z}(w)\;{\rm d}w$, the number passing level $z$ with velocities $w$,
$w+{\rm d}w$ per unit surface per unit time.
\noindent
We assume symmetry with respect to the galactic plane, hence the model
deals only with positive values of $z$ and $w$. 

There is a subset of the pulsar population moving toward the plane after 
having reached maximum distance $z$. Table \ref{table:table1} gives this 
maximum distance, $z_{\rm max}$ for various values of $w_{0}$, and the 
quarter oscillation times $T_{1/4}$, computed with the adopted force field 
described in section 3.1. 
Objects with $w_{0}$ below about 52 km/sec remain during their full 
lifetime below $z=1$ kpc and hence, within the volume of space we study.
For $S=200$ km/sec these represent about one quarter of the population.
For $T=50$ Myr, roughly one half of them will be observed also after they
have reached maximum distance from the plane. Objects with $w_{0}$ in
excess of 50 km/sec will orbit beyond $z=1$ kpc and re-enter the lower
domain on their way back. In these upper parts of their orbits they dwell
relatively long: for $w_{0}=60$ km/sec they re-enter only at the age of
about 50 Myr, for larger $w_{0}$ they dwell even longer beyond 1 kpc.
Hence, it is only in the interval of $w_{0}$ between 52 and 60 km/sec that
we expect to find objects younger than 50 Myr to re-appear in the domain
below $z=1$ kpc. Summarizing, for $T=50$ Myr, a small fraction of the
observed population, about 15 percent, will be on their way back to the
plane. For lower values of $T$, this percentage is accordingly lower. The
simulations take their presence into account.

\begin{table}
\begin{center}
\begin{tabular}{c||c|c|c|c|c|c|c|c|c|c} \hline\hline
${\bf w_{0}}$        & 10  & 20  & 30 & 40 & 50 & 60 & 70 & 80 & 90 & 100 \\
(km/sec)             &     &     &    &    &    &    &    &    &    &     \\ \hline
${\bf z_{\rm max}}$  & 0.12 & 0.27 & 0.45 & 0.68 & 0.95 & 1.22 & 1.56 & 1.90 & 2.21 & 2.6 \\
(kpc)                &     &     &    &    &    &    &    &    &    &     \\ \hline
${\bf T_{\rm 1/4}}$  & 18 & 21   & 25 & 28 & 31.5 & 35 & 38 & 40 & 42 & 43.5  \\
(Myr)                &    &      &    &    &    &    &    &    &    &     \\ \hline\hline
\end{tabular}
\caption{The maximum height from the galactic plane ($z_{\rm max}$) and 
the quarter oscillation time ($T_{\rm 1/4}$) for various initial 
$z$--velocities ($w_{0}$).}
\label{table:table1}
\end{center}
\end{table}

\subsection{Velocities perpendicular to the plane}
The space density of neutron stars at $z=0$ with $w_{0}$,
$w_{0}+{\rm d}w_0$ is 
\begin{equation}
D_{0}(w_{0})\;{\rm d}w_0 = (1/w_{0}) G_{0}(w_{0})\;{\rm
d}w_0 = (C/w_{0})\;{\rm d}w_0.
\end{equation}
\noindent
Since
\begin{equation}
w^{2} = w_{0}^{2} - 2\int_{0}^{z} K(z)\;{\rm d}z
\label{eq:wsqeq}
\end{equation}
\noindent
for objects to reach level $z$, a minimum value of $w_{0}$, denoted by
$w_{\rm min}$ is required:
\begin{eqnarray}
w_{\rm min}^{2} &=& 2\int_{0}^{z} K(z)\;{\rm d}z = 2 P(z) \nonumber \\ 
{\rm Where,} & & P(z) = \int_{0}^{z} K(z)\;{\rm d}z.
\end{eqnarray}
The range of values $w_{0}^{2}$ for objects reaching level $z$ and
beyond is:
\begin{equation}
{\rm Range\;\;in\;\;}w_{0}^{2} = S^{2} - w_{\rm min}^{2}
\end{equation}
This is also the range of the corresponding values $w^{2}$ at level $z$, 
since $w_{\rm min}$ becomes zero and $S$ becomes $w_{s}$,
\begin{equation}
{\rm where\;\;} w_{s}^{2} = S^{2} - 2\int_{0}^{z} K(z)\;{\rm d}z = S^{2} -
2P_{z} = S^{2} - w_{\rm min}^{2}
\end{equation}
However, the range of $w$ at level $z$, {\it i.e.,} from zero to $w_{s}$,
is broader than the corresponding range at birth in the ratio
$w_{s}/(S-w_{m})$. For instance, if we assume S=180 km/sec, for reaching
level 420 pc, $w_{0}$ should range from 30 to 180 km/sec, {\it i.e.,}
150 km/sec, whereas the resulting range at level $z$=420 pc becomes $w=0$
to $w_{s} = 177$ km/sec. 

Since for given $z$, from equation~\ref{eq:wsqeq}, $\Delta w^{2} =
\Delta_{0}(w_{0}^{2})$, we have, $w_{0}\;\Delta w_{0} =
w\;\Delta w$. Hence, the broadening for given $w$ is in the ratio
$w_{0}/w$. 

Accordingly, whereas $G_{0}(w_{0})\;{\rm d}w$ is flat, for a given
level $z$, $G_{z}(w)\;{\rm d}w$ is proportional to $w/w_{0}$, with
$w$ and $w_{0}$ related by equation~\ref{eq:wsqeq}.

The number of objects per unit volume at level $z$ with $w$ in the interval
$w$, $w+{\rm d}w$ will be,
\begin{equation}
D_{z}(w)\;{\rm d}w = (1/w) G_{z}(w)\;{\rm d}w.
\end{equation}
\noindent
Due to the broadening just mentioned,
\begin{equation}
G_{z}(w) = (w/w_{0}) G_{0}(w_{0})
\end{equation}
\noindent
Hence, $D_{z}(w) = (C/w_{0})$, and we have the well known
relation, $D_{z}(w) = D_{0}(w_{0})$ for pairs of values $w$ and
$w_{0}$ related by equation~\ref{eq:wsqeq}.

The above relations do not take into account the disappearance from the
sample of the pulsars older than age $T$. We denote by $w_{t}$ the minimum
velocity at $z=0$ required for reaching level $z$ within the time-span $T$.
$w_{t}$ is defined by:

\begin{equation}
T\;=\;\int_{0}^{z}\frac{{\rm d}z}{w(z)}\;=\;\int_{0}^{z} \frac{{\rm
d}z}{\sqrt{w_{t}^{2}-2P}}.
\end{equation}

\noindent
Thus, there is an interval of velocities at level $z$ from zero to $w_{z}$,
in which no objects will occur. Again, we have,
\[
w_{z}^{2}\;=\; w_{t}^{2} - 2P
\]

Accordingly, at level $z$, the distribution of velocities $w$ can be
described by the above $D_{z}(w)\;{\rm d}w$ truncated at the low velocity end
by eliminating the interval $w=0$ to $w=w_{z}$.

Figures~\ref{fig:zslabvel} demonstrate this effect. They have been obtained
numerically for an adopted value $S=180$ km/sec and $T=8$ Myr. The five 
histograms correspond to five different $z$-slabs (between 0 and 1 kpc), 
each of width 200 pc. In the highest layer we note the absence of pulsars 
in the velocity range $w=0$ to $90$ km/sec. At lower $z$-levels the empty 
interval is accordingly smaller. At the high-velocity end the limiting 
velocity $w_l$ is fixed by $w_l^2 = (S^2 - 2P)$. The fact that the cut-off 
is not sharp is due to the combination of all $z$-values within the slab. 

\subsection{Velocities parallel to the plane}
\label{sec-vpareq}
\noindent
We denote by, \\
\noindent
$v$, the component of the velocity parallel to the galactic plane 
corresponding to the proper motion in galactic longitude: 
\[
v = d\,\mu_{l}\;\cos b
\]
\noindent
Since the distribution of $v$ is symmetric we consider only the absolute
value of $v$. At level $z$, for a given $T$, $v$ ranges from 0 to $v_{l} =
(S^{2} - w_{t}^{2})^{1/2}$. The predicted distribution will be shown to be:

\begin{equation}
D(v)\;{\rm d}v\;=\; C^{'}\;S\left\{\pi/2 - \arcsin \left[ \frac{w_{t}}
{\sqrt{S^{2}-v^{2}}}\right]\right\}\;{\rm d}v
\label{eq:dvdv}
\end{equation}

\noindent
where $C^{'}$ is a constant defined below.

We define by $N(v,w)$ the number of objects at level $z$ per unit of $v$ and
of $w$. Equation~\ref{eq:dvdv} then follows from

\[
D(v)\;{\rm d}v\;=\;{\rm d}v\int_{w_{t}}^{w_{l}} N(v,w)\;{\rm d}w\;=\; {\rm
d}v\int_{w_{t}}^{w_{l}}N(0,w)\;\frac{R\;{\rm d}w}{\sqrt{R^{2}-v^{2}}}
\]

\noindent
where, $R^{2} = (S^{2} - w^{2})$, $w_{l}^{2} = (S^{2} - v^{2})$, and $N(0,w)
= N(0,0)\;S/\sqrt{S^{2}-v^{2}}$, so that,

\begin{eqnarray}
D(v)\;{\rm d}v &=& {\rm d}v\; S\; N(0,0)\int_{w_{t}}^{w_{l}}\frac{{\rm
d}w}{\sqrt{S^{2}-v^{2}-w^{2}}} \nonumber \\
 &=& S\;N(0,0)\left[ \pi/2 -
\arcsin\frac{w_{t}}{\sqrt{S^{2}-v^{2}}}\right]\;{\rm d}v. \nonumber
\end{eqnarray}

\noindent
where $C^{'} = N(0,0)$ is the number per unit $v$ and unit $w$ at $v=0$, 
$w=0$ at $z=0$.

For low $z$-levels the distribution $D(v)\;{\rm d}v$ is nearly flat up to
$v$ close to $v_{l}$. This is due to the fact that the second term in 
equation~\ref{eq:dvdv} becomes significant with respect to the first,
constant term, only at high $v$. For instance, for $S=200$ km/sec and
$T=20$ Myr, at $v=0.9v_{l}$, the ratio between the second and the first
term near $z=0$ is only 0.036. For high $z$-levels the distribution becomes 
less flat. Figure~\ref{fig:vlsim} demonstrates this effect. As we chose
higher values of $T$, the nature of the curves for higher $z$ becomes more
similar to the one for the lowest $z$. The truncated nature of $D(v)$, 
particularly at low $z$, provides an important tool in estimating $S$; 
the extreme observed value, $v_{l}$ is an approximate measure 
of $S$ virtually independent of the assumed life time $T$.

\setcounter{figure}{2}

\begin{figure}
\epsfig{file=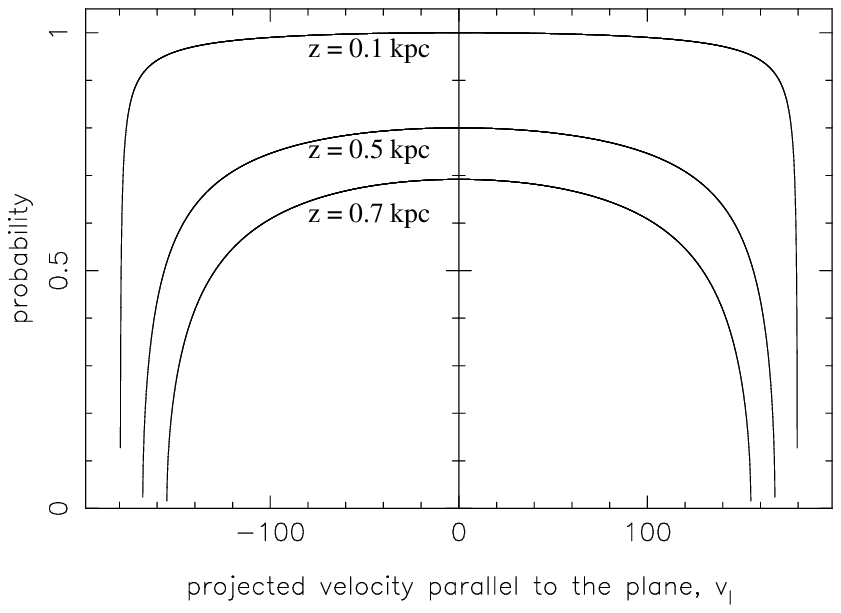,height=7.5cm}
\caption[]{Distribution of velocities $D(v)$ parallel to the plane, as given
by equation \ref{eq:dvdv}. The three curves correspond to $z$ = 0.1, 0.5 \& 
0.7 kpc, respectively, for $S=180$ km/sec and $T=8$ Myr.}
\label{fig:vlsim}
\end{figure}

\section{Simulations}
\label{sec-simulations}
We did a Monte Carlo simulation to generate a large number of objects in
the solar neighbourhood (with initial value of $z$=0), keeping the
surface density constant. As mentioned in the previous sections, we
assumed a simplistic model with all neutron stars being ejected out
isotropically with a constant speed ($S$). This assumption corresponds to
having initial velocities in the $z$--direction ranging from zero to $S$
with uniform probability. We evolve them in an assumed local
gravitational potential to get their velocity and spatial distributions,
and in addition we studied the distributions of their characteristic ages.
Each object was assigned an age chosen randomly with uniform
probability between zero and $T$. Given the distribution of characteristic
ages of known pulsars, we chose a value of $T=50$ Myr as an optimal choice
in order to include a sufficiently large sample, and at the same time
describe the kinematics with our potential model. The presence of the 
objects which turn over after reaching their maximum $z$--amplitude of 
oscillation within their age are accounted for.

\subsection{Local gravitational potential}
To study the kinematics of the generated objects we assumed the local 
gravitational potential function given by Kuijken \& Gilmore (1989). The
vertical acceleration due to this potential is given by (Bhattacharya {\it
et al.})

\begin{equation}
g_{z}\;=\;1.04\times 10^{-3}\;\left[\frac{1.26\;z} {\sqrt{z^{2}+0.18^{2}}}
+ 0.58\;z\right]
\end{equation}

\noindent
where $g_{z}$ is in units of ${\rm kpc/Myr^{2}}$ and $z$ in kpc. We use
this function to evolve every one of the generated objects from their
initial position and velocity to their current value after a time $t$. 

\subsection{Further assumptions for simulation}
\label{sec-assim}
We assumed a Gaussian distribution of fields 
in logarithmic scale, with $<\log B(G)>=12.2$ with a root mean square spread
around the mean of 0.3. The choice of initial periods will be discussed in 
section \ref{sec-agep}. With an assumption that the magnetic field does 
not decay significantly within the time scales of our interest ($\sim$50 
Myr) the final rotational period of the pulsar can be easily calculated 
with the simple dipole formula 

\begin{equation}
B^{2}\;=\;\frac{3Ic^{3}}{8\pi R^{6}}P\dot{P}
\label{eq:dipole}
\end{equation}

\noindent
where $c$ is the velocity of light, $I$ is the moment of inertia of the
neutron star, and $\dot{P}$ is the time derivative of the rotation period
$P$.

The simulation was repeated for various values of $S$ (ranging from 90
km/sec to 350 km/sec) and $T$ (from 15 Myr to 50 Myr). At the end of the
evolution, pulsars which have the quantity $(B/P^{2})<2\times 10^{11}$ were
neglected from the simulation, since it is believed that the pulsar
activity ceases to continue below this limit.

\subsection{Selection filter}
\label{sec-selection}
After evolving the generated objects in the galactic potential, their final
position, velocity components, rotation period and characteristic age are
noted down. After compensating for observational selection effects a subset 
of this population was selected to compare with the known sample. The
procedure we have adopted for this task is the same as the one described by
Deshpande {\it et al.} (1995). We selected only those pulsars which are, in 
principle, detected by any one of the eight major pulsar surveys considered, 
namely (1) Jodrell Bank survey, (2) U. Mass--Arecibo survey, (3) Second 
Molonglo survey, (4) U. Mass--NRAO survey, (5) Princeton--NRAO Phase I
survey, (6) Princeton--NRAO Phase II survey, (7) Princeton--Arecibo survey,
and (8) Jodrell Bank--1400 MHz survey. Since our analysis deals with only 
the solar neighbourhood (projected distance onto the plane less than 2 kpc,
and $|z|<1$ kpc) 
the effect due to interstellar scattering and dispersion are ignored. While
comparing the results of the simulation with the known sample, only the 
subset of the simulated sample ({\it detectable} simulated sample) was
used.

\begin{figure}
\epsfig{file=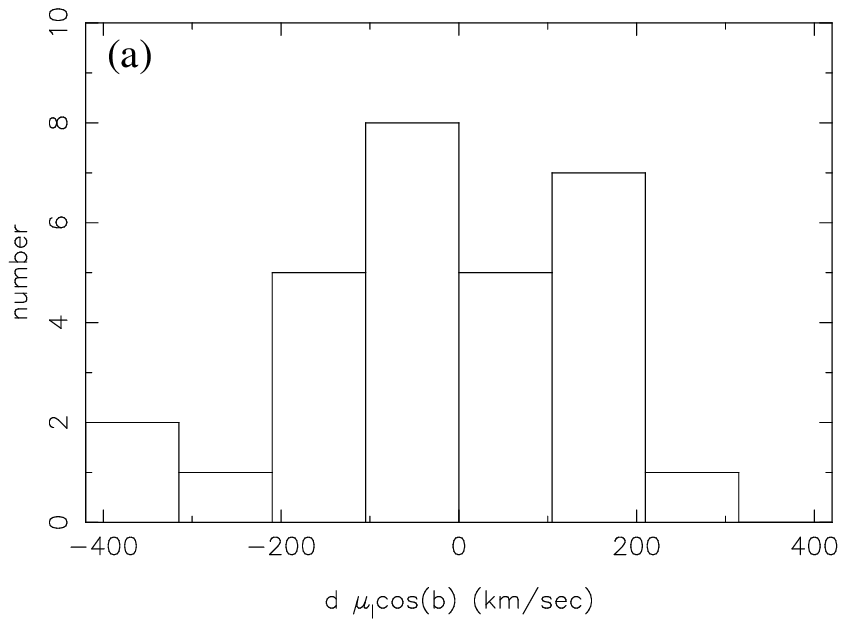,height=7cm}
\vspace*{1cm}
\epsfig{file=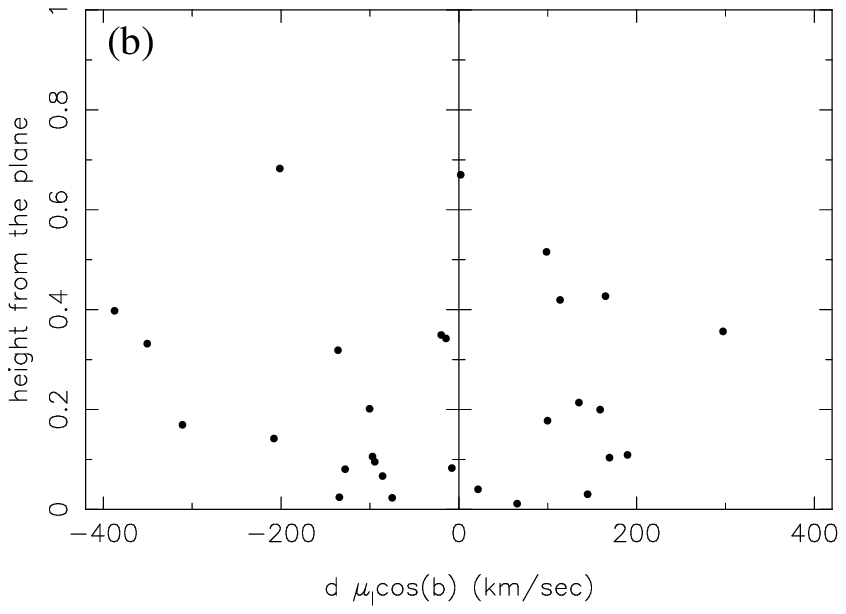,height=7cm}
\caption[]{Velocities parallel to the plane corresponding to the proper 
motions in galactic longitude. ($a$) Distribution
of these velocities, ($b$) Distribution of velocities as a function of
height from the galactic plane. Only those pulsars whose proper motion
errors ($1\sigma$) are less than 94 km/sec (20 A.U./yr) are considered
for this figure.}
\label{fig:muldz}
\end{figure}

For consistency, even the sample of known pulsars were passed through this
filter to select a subsample, which was finally used for comparison with
the simulated sample.

\section{Comparison with observations}
\subsection{Velocities parallel to the plane}
As the first step, we consider the distribution of velocity components
parallel to the plane, $D(v)$, because as it was explained in section 
\ref{sec-vpareq}, these provide the estimate of $S$. From the compilation 
of Taylor {\it et al.} (1995) we use  those proper motion measurements for 
which the quantity $d\times\sigma<94$ km/sec (20 A.U./yr), where $\sigma$ is 
the formal error in the proper motion quoted by those authors. We do not 
apply corrections for solar motion and differential galactic rotation, since 
within the range of distances considered, these are negligible as compared 
to the true proper motions and their observational errors. Figure 
\ref{fig:muldz}(a) gives the distribution of these velocities, and 
figure \ref{fig:muldz}(b) gives their distribution as a function of the 
height from the galactic plane. We note that with the exception of 5 
objects, all are confined within $\pm 210$ km/sec. In comparison with figure 
\ref{fig:vlsim} we may conclude that the value of the unique velocity must 
be roughly around about 200 km/sec. Clearly, values of $S$ in excess of 250
km/sec are incompatible with the proper motions.

\subsection{Density Distribution perpendicular to the plane}
\label{sec-zdist}

\begin{figure}
\epsfig{file=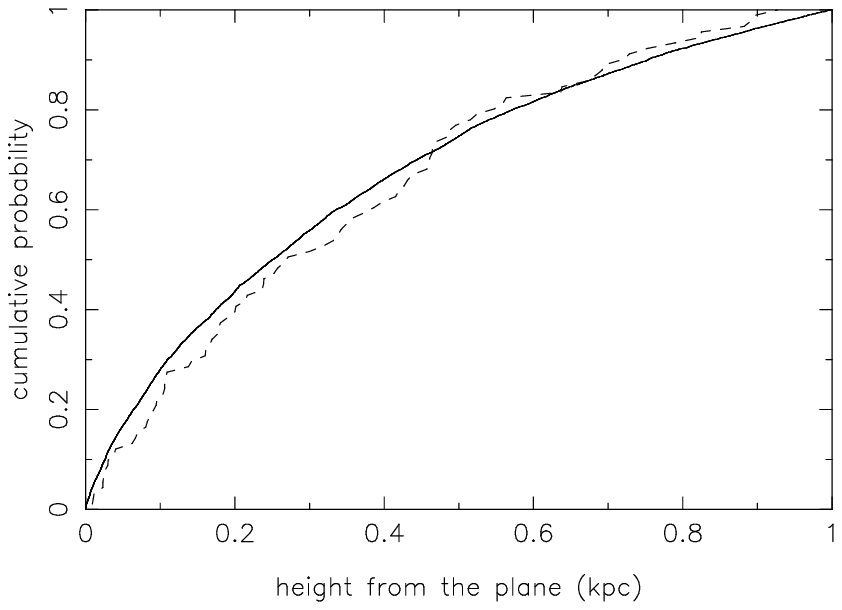,height=7.5cm}
\caption[]{The cumulative number distribution of pulsars as
a function of height from the galactic plane. Dash line is for the known
sample and the solid line is the best fit corresponding to $S=180$ km/sec
and $T=50$ Myr.}
\label{fig:zmatch}
\end{figure}

Hartman \& Verbunt (1994), from their study of evolution of neutron stars
in various models of the local gravitational potential, demonstrate the
effect of observational selection for understanding the distribution of
pulsars perpendicular to the plane. Since observations tend to detect more
luminous pulsars (which are younger and closer to the plane), irrespective
of the nature of the {\it true} distribution the {\it observable}
distribution is roughly the same. This has been seen in our simulations
too. While comparing the simulated sample (after compensating for the 
selection effects) with the observed, a satisfactory Kolmogorov -- Smirnov 
probability was achieved for many different combinations of $S$ and $T$. For 
$S=180$ km/sec and $T=50$ Myr (figure \ref{fig:zmatch}) the K--S probability 
was 59.6\%, and for many other combinations of $(S,T)$, like (250,50), 
(350,50) the K--S probability was roughly as high as for (180,50). 
Therefore, one can conclude that the density distribution perpendicular to 
the galactic plane alone is not a severe constraint on the kinematic 
properties of pulsars, and one needs to study other properties like the 
velocity components and the distribution of characteristic ages, etc.

\subsection{Velocities perpendicular to the plane}
\begin{figure}
\epsfig{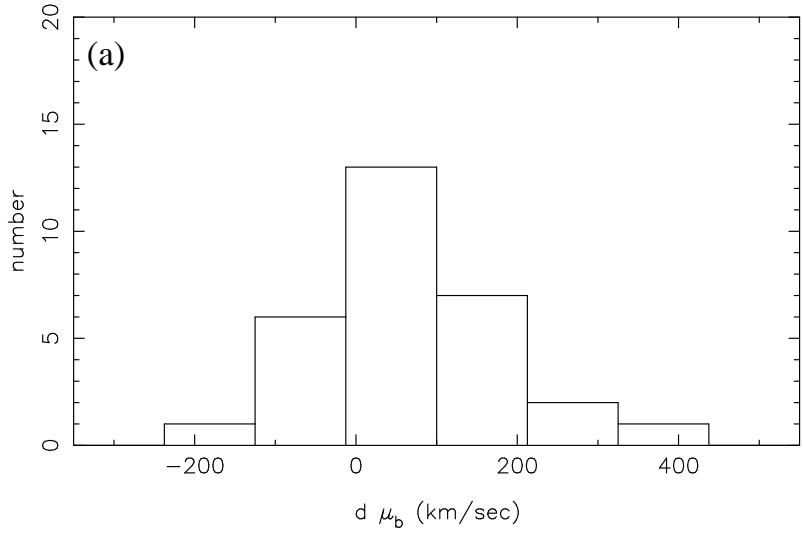}
\vspace*{1cm}
\epsfig{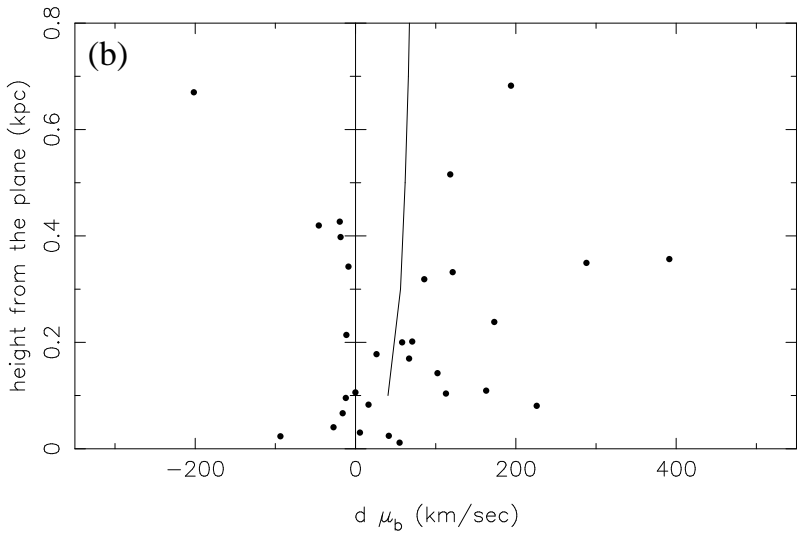}
\caption[]{($a$) Distribution of $d\times\mu_{b}$ of known pulsars, ($b$)
distribution of known pulsars as a function of $d\times\mu_{b}$ and the
height from the plane. Only those pulsars with $|b|<50^{\circ}$ and velocity 
error less than 94 km/sec are included. The solid line gives the mean of the 
simulated sample for $S=180$ km/sec.,$T=50$ Myr.}
\label{fig:mubdz}
\end{figure}

We do not have measured $w$, since there is no way of measuring
the radial component of the velocity vector. However, for pulsars with
sufficiently low galactic latitude (say, $|b|<50^{\circ}$), we can
approximate $w=\mu_{b}\times d$ provided $\cos b$ is close to unity. Out 
of the 30 pulsars selected, 23 have 
$\cos b>0.87$, 5 have $\cos b = 0.78-0.87$, and 2 have $\cos b = 
0.72-0.78$. Figure \ref{fig:mubdz}(a) gives the distribution of $(\mu_{b}
\times d)$. Figure \ref{fig:mubdz}(b) shows the distribution of $\mu_{b}
\times d$ against the height from the galactic plane. Objects at negative 
$z$ are plotted with reversed $z$ and reversed $\mu_{b}\times d$. The 
asymmetry in the distribution of $\mu_{b}\times d$ in figure 
\ref{fig:mubdz}(a) shows the flow of objects away from the galactic plane. 

Once a considerably richer sample of accurate pulsar proper motions is 
available, it will be of great interest to make the detailed comparison
between the observed distribution and the theoretical one of figure 
\ref{fig:zslabvel}, and check its rapidly varying shape with increasing
$z$. The predicted run of the mean velocity $v$ with increasing $z$ is in
harmony with observations, as shown in figure \ref{fig:mubdz}(b).

\subsection{Distribution of characteristic ages}
\label{sec-agep}
As described in section \ref{sec-assim} we have assumed that the magnetic
fields at birth are distributed as a Gaussian in $\log B$ around 12.2 with
an r.m.s. of 0.3. Since many of the earlier works (Vivekanand \& Narayan 1981;
Srinivasan {\it et al.} 1984; Narayan 1987; Narayan \& Ostriker 1990; Deshpande
{\it et al.} 1995) suggest that pulsars are born with rotation periods as long
as a few hundred milliseconds, for the simulation, all pulsars were assumed to 
be born with rotation period of 0.1 sec. With the assumption that the magnetic 
field does not decay with time, the period was evolved with the simple dipole
formula \ref{eq:dipole}. Objects were evolved in the
local gravitational potential for their corresponding ages (distributed between
zero and $T$ with uniform probability). At the end of
the simulation pulsars with $B/P^{2} < 2\times 10^{11}$ Gs$^{-2}$ were
neglected, since the pulsar activity is believed to cease below this limit.
After applying the {\it selection effect filter}, a subset of the simulated
sample (detectable subset of the simulated sample) was selected for
comparison with the known sample.

\begin{figure}
\epsfig{file=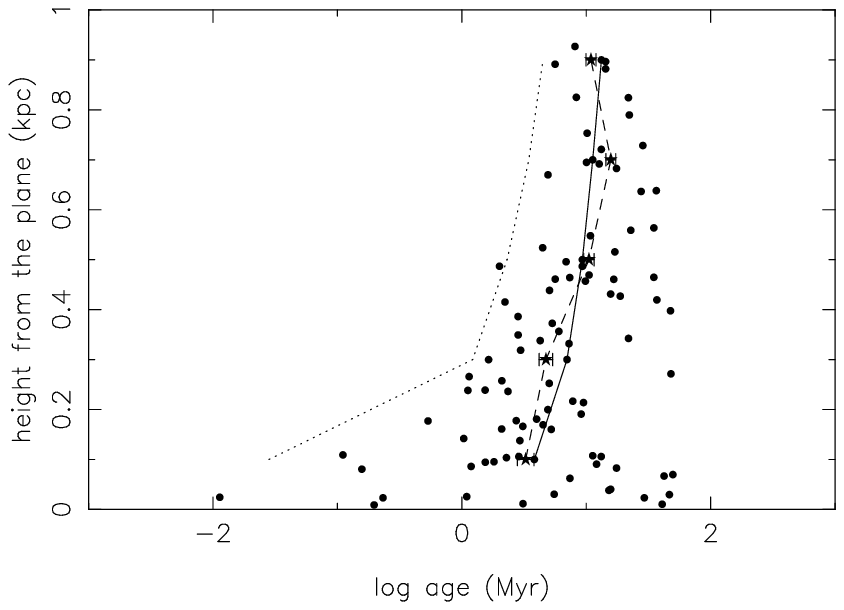,height=7cm}
\caption[]{Distribution of known pulsars in $z-\log\tau_{\rm ch}$ plane. 
The dash line indicates the average $\log\tau_{\rm ch}$ as a function of 
$z$, for the known sample. Solid line indicates the corresponding quantity 
for the simulated sample for $S$=180 km/sec and $T$=50 Myr. The dotted 
lines give the predicted lower bound for $\tau_{\rm ch}$ as a function of 
$z$.The error bars indicate the uncertainty in the mean value.}
\label{fig:ztaup}
\end{figure}

Figure \ref{fig:ztaup} gives the distribution of known pulsars in the
$z-\log\tau_{\rm ch}$ plane (where $\tau_{\rm ch}$ is the characteristic
age of the pulsar). The `dash' line gives the observed mean value of $\log 
\tau_{\rm ch}$ in the five different $z$--slabs from zero to 1 kpc, and the 
solid line gives the corresponding mean value of the ``detectable'' 
simulated sample for the combination of $S=180$ km/sec and $T=50$ Myr. The 
``dotted'' line gives the possible lower limit of $\tau_{\rm ch}$ for a 
given height from the plane. {\it i.e.}, with initial velocities, $w_{0}$, 
ranging from zero to $S$, the object can reach a given height $z$, only if 
its age is greater than a certain value.

As one can see from this figure, our model fits with the 
observed distribution of $\log\tau_{\rm ch}$ as a function of $z$, quite 
satisfactorily. The fact that the envelope in figure \ref{fig:ztaup} (given 
by the dotted line) encloses all the pulsars means that almost all pulsars 
are born close to the plane and they migrate by the virtue of their 
velocities. As it turns out, to constrain the combination of $S$ and $T$, 
the distribution of $\log\tau_{\rm ch}$ can be used effectively.

Ideally we would have liked to compare the distribution of $\log\tau_{\rm ch}$ 
in each $z$--slab with the corresponding observed distribution. 
However, small number of the observed sample prevents us from doing so.

\section{Conclusions}
We have explored the hypothesis that pulsars are born in the galactic plane 
with a unique kick velocity $S$, distributed isotropically. We therefore 
investigate the pulsars in the solar neighbourhood by studying the
distributions of the velocity components parallel and perpendicular to the
galactic plane, and the characteristic age distribution. We find that the 
known sample of pulsars in the solar neighbourhood is consistent with kick 
velocities of about 200 km/sec.  

We wish to emphasis that our analysis of the nearby pulsars leaves no room
for the existence of an appreciable fraction of pulsars with velocities 
much higher than 200 km/sec. These should have shown up in the plots of
figures \ref{fig:muldz} \& \ref{fig:mubdz}. Evidence for very high velocities 
came from a few reliable proper motion measurements (Bailes {\it et al.} 
1990; Harrison {\it et al.} 1993; Fomalont {\it et al.} 1992) and velocities 
inferred from the association of a few pulsars with supernova remnants 
(Caraveo 1993; Frail, Goss \& Whiteoak 1994). The reliability of such 
associations has been questioned by Kaspi(1996).

The analysis by Brandt \& Podsiadlowski (1995) shows that the assumed kick 
speeds of the order of 450 km/sec are inconsistent with the eccentricity --
orbital period distribution of binaries which contain neutron stars, and
it is consistent with kick speeds of 200 km/sec. This supports our result.

The velocity distribution derived by Lyne \& Lorimer (1993) and Hansen \&
Phinney (1997), referred to in the introduction and figure 1, show a
considerable fraction of pulsars with velocities around 200 km/sec, the
value of $S$ we arrived at. They differ from our model in two respects:
\begin{enumerate}
\item Both of them show maximum density in velocity space around velocity
zero, whereas our model gives zero density at zero velocity. 
\item Both of them show a long tail towards the high-speed end. This is
recognised in our plots of the velocity components $v$ and $w$ (figures 4
\& 6), but this represents a small fraction only of the total distribution. 
\end{enumerate}

The recent paper by Fomalont {\it et al.} (1997) has brought out improved
proper motion measurements for about 15 pulsars. Among those pulsars with
distance less than 2 kpc and galactic latitude less than 50 degrees,
on the basis of the earlier measurements we had ignored 1556--44 \& 
1749--28, since they had errors greater than 20 A.U./yr. For 1919+21 \&
2045--16 the improved proper motion measurements seem to be quite
different from the earlier measurements. However, since these changes are
unlikely to change the results in this paper we have done our analysis on
the basis of the earlier measurements.

To summarise the main result, we show that if pulsars are postulated to 
derive a unique isotropic supernova kick speed at their birth around 200 
km/sec, the distribution of the proper motions and the characteristic 
ages of pulsars in the solar neighbourhood can then be understood 
satisfactorily.

\section*{Acknowledgement}
We would like to thank Dipankar Bhattacharya for his very valuable comments 
which have helped us a great deal to improve the manuscript. 

\section*{References}
\begin{enumerate}
\item Bailes M., Manchester R.N., Kasteven M.J., Norris R.P. \& Reynolds 
      J.E. 1990, Monthly Notices Roy. Astr. Soc., 247, 322
\item Bhattacharya D., Wijers R.A.M.J., Hartman J.W. \& Verbunt F. 1991, 
      Astron. Astrophysics, 254, 198
\item Blaauw A. 1961, Bull. Astr. Inst. Netherlands, 15, 265
\item Brandt N. \& Podsiadlowski P. 1995, Monthly Notices Roy. Astr. Soc.,
      274, 461
\item Caraveo P.A. 1993, Astrophys. J., 415, L111
\item Deshpande A.A., Ramachandran R. \& Srinivasan G. 1995, J. Astrophys.
      Astr., 16, 53
\item Fomalont E.B., Goss W.M., Lyne A.G., Manchester R.N. \& Justtanont K. 
      1992, Monthly Notices Roy. Astr. Soc., 258, 497
\item Fomalont E.B., Goss W.M., Manchester R.N. \& Lyne A.G. 1997, Monthly 
      Notices Roy. Astr. Soc., 286, 81
\item Frail D.A., Goss W.M. \& Whiteoak J.B.Z. 1994, Astrophys. J., 437, 781
\item Gott J.R., Gunn J.E. \& Ostriker J.P. 1970, Astrophys. J., 160, L91
\item Hansen B.M.S. \& Phinney E.S. 1997, Monthly Notices Roy. Astr. Soc., 
      291, 569
\item Harrison P.A., Lyne A.G. \& Anderson B. 1993, Monthly Notices Roy. 
      Astr. Soc., 261, 113
\item Harrison E.R. \& Tadimaru E. 1975, Astrophys. J., 201, 447
\item Hartman J.W. 1997, Ph.D. thesis, Utrecht University
\item Kaspi V.M. 1996, in Johnston S., Walker M.A., Bailes M. eds., ASP Conf. 
      Ser. Vol. 105, Pulsars: Problems \& Progress. Astron. Soc. Pac. 
      San Francisco, P.375
\item Kuijken K. \& Gilmore G. 1987, Monthly Notices Roy. Astr. Soc., 
      239, 571
\item Lyne A.G. \& Lorimer D.R. 1993, Nature, 369, 127
\item Lyne A.G., Anderson B. \& Salter M.J. 1982, Monthly Notices Roy. 
      Astr. Soc., 201, 503
\item Narayan R. 1987, Astrophys. J., 319, 162.
\item Narayan R. \& Ostriker J.P. 1990, Astrophys. J., 352, 222.
\item Shklovskii I.S. 1970, Astr. Zu., 46, 715
\item Srinivasan G., Bhattacharya D. \& Dwarakanath K.S. 1984, J. Astrophys.
      Astr., 5, 403
\item Taylor J.H., Manchester R.N. \& Lyne A.G. 1993, Astrophys. J. Supp., 
      88, 529
\item Vivekakand M. \& Narayan R. 1981, J. Astrophys. Astr., 2, 315.
\end{enumerate}

\setcounter{figure}{1}
\begin{figure}
\epsfig{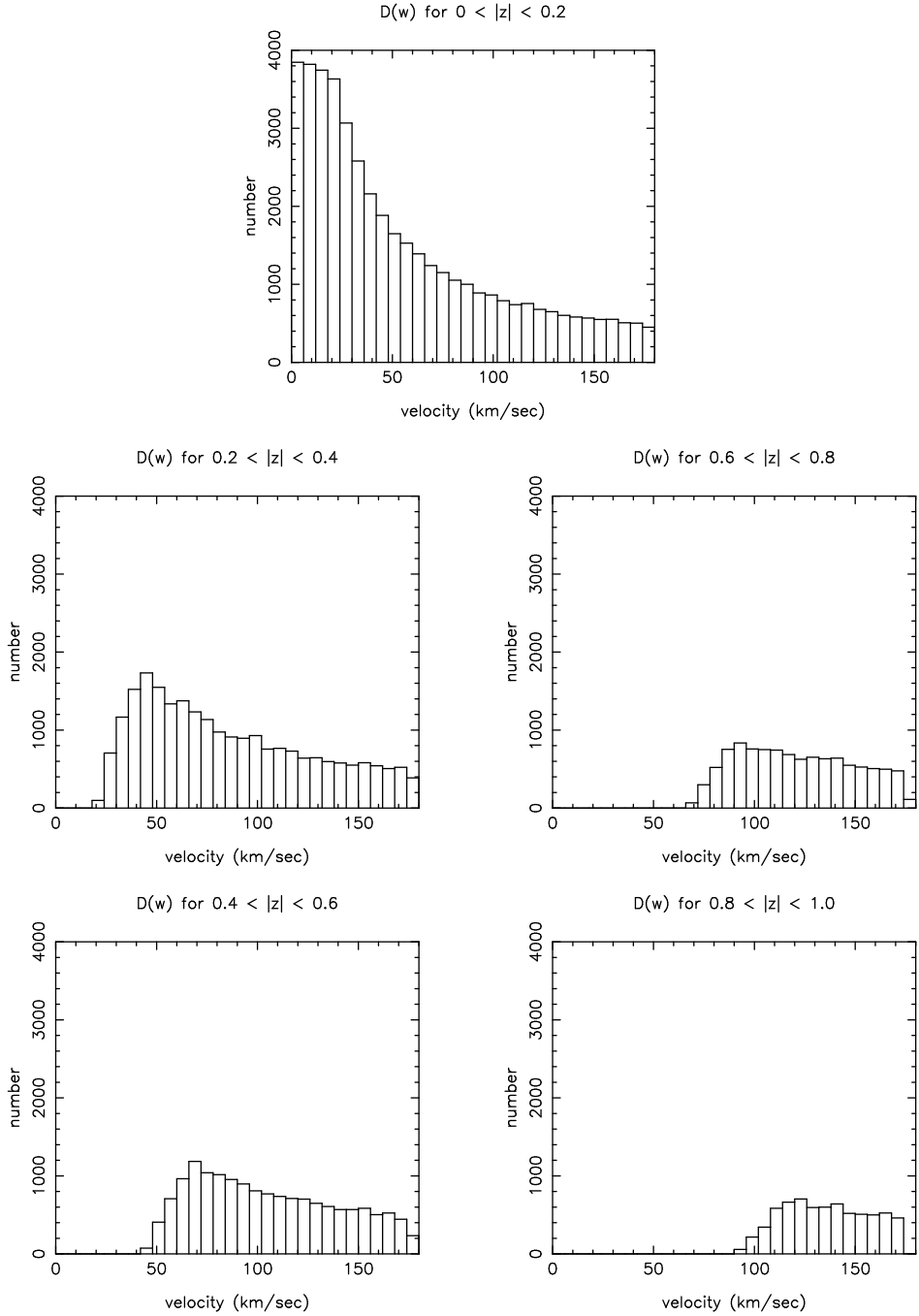}
\caption[]{Distribution of velocities perpendicular to the plane of 
pulsars in different $z$--slabs after evolving in the potential for ages 
up to $T$. This plot corresponds to $S=180$ km/sec and $T=8$ Myr.}
\label{fig:zslabvel}
\end{figure}

\end{document}